\documentclass{article}
\usepackage{graphicx}
\newcommand{\half}{\frac{1}{2}}
\newcommand{\bqu}{{\bf q}}
\begin{document}
\begin{center}
{\Large \bf A MODEL FOR TWO-PROTON EMISSION} \\
\vspace {0.2 cm}
{\Large \bf  INDUCED BY ELECTRON SCATTERING} \\
\vspace {1. cm}
{\Large  Marta Anguiano}
\hspace{0.2cm} and \hspace{0.2cm} {\Large Giampaolo Co'} \\ 
\vspace{0.3 cm}
{Dipartimento di Fisica,  Universit\`a di Lecce, \\
Istituto Nazionale di Fisica Nucleare  sez. di Lecce, 
\\ I-73100 Lecce, Italy} \\
\vspace{.5cm}
and \\
\vspace{.5cm}
{\Large Antonio M. Lallena} \\
\vspace{.3cm}
{ Departamento de F\'{\i}sica
Moderna, Universidad de Granada, \\
E-18071 Granada, Spain} \\
\end{center}
\vspace {1. cm} 
\begin{abstract}
  We present a model to describe the emission of two protons
  in electron scattering experiments.  The process is induced by
  one-body electromagnetic operators acting together with short-range
  correlations, and by two-body $\Delta$ currents.  The model includes
  all the diagrams containing a single correlation function.  The
  sensitivity of the cross section to the details of the correlation
  function is studied by using realistic and schematic correlations.
  Results for the $^{16}$O nucleus are presented.
\end{abstract}

\vspace {1. cm}
This work belongs to a series dedicated to the study of the effects of
the short-range correlations (SRC) on electromagnetically induced
nuclear excitations. We have studied inclusive electron scattering
cross sections in both the excitation of discrete low-lying states
\cite{mok00} and in the quasi-elastic peak \cite{co01}, and also
one-nucleon emission induced by real photons \cite{ang02} and
electrons \cite{mok01}.

In all these calculations the SRC effects were hidden by the large
contribution coming from the uncorrelated one-body electromagnetic
operators. A possibility of eliminating these contributions is to
investigate processes where two-nucleons are emitted. We have studied
(e,e'2p) processes where, in coincidence with the scattered electron,
also two protons are detected.
In this contribution we shall present cross sections as a function of
the detection angle of one of the emitted protons, all the other
kinematics variables being fixed. 

In our model we describe the nuclear many-body states as:
\begin{equation}
|\widetilde{\Psi}_n \rangle \, = \,
\frac {|\Psi_n \rangle } {\langle \Psi_n |\Psi_n\rangle ^{\half}}
\,\, ,
\,\,\,\,\,\,\,
|\Psi_n\rangle \,= F \,|\Phi_n\rangle
\,\, ,
\,\,\,\,\,\,\,
F = \prod_{i<j} f(r_{ij}) \,\, ,
\end{equation}
where $|\Phi_n \rangle$ is a Slater determinant and $F$
is a correlation function we assumed
formed by a product of two-body scalar correlation functions $f(r)$.
The evaluation of the cross section 
\begin{equation}
\nonumber
d \sigma \, \sim \, \sum_n \, |\langle \widetilde{\Psi}_n |\,
 O(\bqu)\, |\widetilde{\Psi}_0 \rangle |^2 \,
\delta (E_n - E_0 - \omega)
\end{equation}
implies the calculation of 
\begin{eqnarray}
\label{eq:xi1}
\xi_n(\bqu)\, &=& \, \frac { \langle \Psi_n | \,O(\bqu)\, |\Psi_0
 \rangle } 
 {\langle \Psi_n |\Psi_n\rangle ^{\half}\,\langle \Psi_0 |\Psi_0\rangle
 ^{\half} } \\
\label{eq:xi2}
 &=&
\frac{ \langle \Phi_n |F^+  
O(\bqu)  F | \Phi_0 \rangle }
{
\langle \Phi_n | F^+ F | \Phi_n \rangle ^\half
\langle \Phi_n | F^+ F | \Phi_n \rangle ^\half} \\
\label{eq:xi3}
 &=&
\langle \Phi_n | O(\bqu) F^2 |\Phi_0 \rangle_L
=  \langle\Phi_n|
O(\bqu)\prod^A_{i<j}(1+h_{ij})|\Phi_o \rangle_L \,\, ,
\end{eqnarray}
where $O(\bqu)$ is the one-body electromagnetic operator inducing the
transition. The denominator in Eq. (\ref{eq:xi2}) allows the
elimination of the unlinked diagrams. This is the meaning of the
subindex $L$ in equation (\ref{eq:xi3}), indicating that only the linked
diagrams should be considered.  Since in our calculation $F$ is a scalar
function, it commutes with $O(\bqu)$, and this allows us to write the
final expression where we have defined $h_{ij} = 1 - f(r_{ij})$.  
\begin{figure}[ht]
\begin{center}
\includegraphics[bb=20 150 600 750,angle=0,scale=0.5]
 {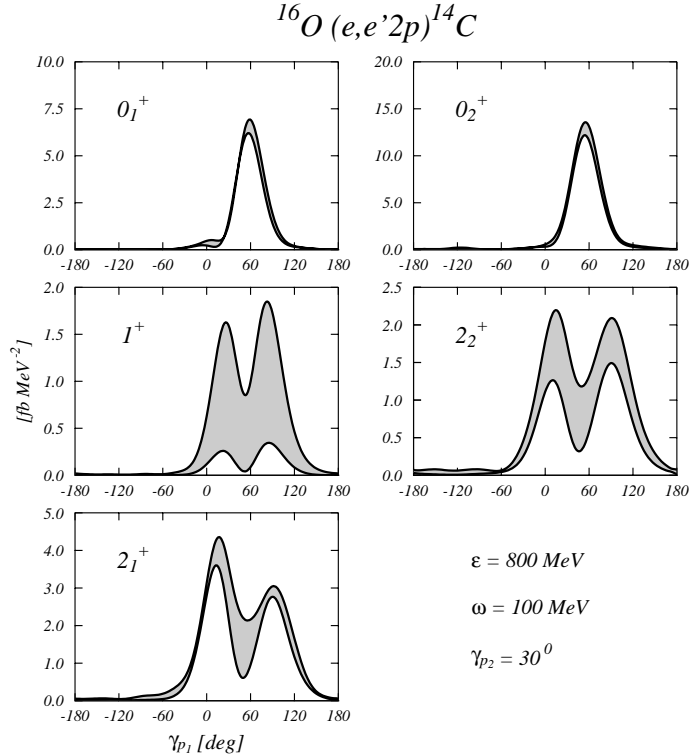}
\end{center}
\caption{
 Cross sections for the $^{16}$O(e,e'2p)$^{14}$C
 process. We indicated with $\epsilon$ the electron incoming energy,
 with $\omega$ the nuclear excitation energy, with $\gamma_{p_{2}}$
 the emission angle of one proton and with $\gamma_{p_{1}}$ the
 emission angle of the second proton. The numbers in the panels
 indicate the quantum numbers of the $^{14}$C residual nucleus. We
 calculated only the final states that can be reached by emitting
 protons from the $p$ levels of the $^{16}$O nucleus. The gray bands
 are produced by the uncertainty on the theoretical inputs of the
 calculations, done with the same correlation function.
 }
\label{fig:minmax}
\end{figure}
This
final writing shows the fact that the transition matrix element can be
written as a sum of power of $h$. The minimum power is zero,
represented by the 1 in Eq. (\ref{eq:xi3}) and leading to the
uncorrelated transition. The maximum power is $A$. In our approach we
use the approximation of restricting the sum to the first power in $h$
\begin{eqnarray}
\label{eq:xi11}
\xi_n(\bqu)  \longrightarrow  \xi_n^1(\bqu) & = &  
\langle \Phi_n | \,O(\bqu)\, 
( 1+ \sum_{i<j}\, h_{ij}  ) \,|\Phi_0  \rangle_L 
\\
\nonumber
\label{eq:xi12}
 &=& 
  \langle \Phi_n (2p 2h)|  
\,O(\bqu)\, \sum_{1<j} h(r_{1,j}) 
\,|\Phi_0  \rangle_L \, \\
\label{eq;xi13}
& + &
 \langle \Phi_n (2p 2h)|  \,O(\bqu)\, 
\sum_{1<i<j}\, h(r_{i,j})
\,|\Phi_0 \rangle_L \,\, .
\end{eqnarray}

In the last expression we show that, when the final state is composed
by two particles in the continuum, the uncorrelated term does not act.
For this reason there are only two types of terms. A first one where
one of the coordinates of the function $h$ is also the coordinate
where the external operator is acting on. This term leads to two-point
diagrams. In the second type of terms, leading to three-point
diagrams, the three coordinates are different. The presence of the two
kinds of diagrams, necessary for the proper normalization of the final
state of the nucleus, is a novelty of our calculations.

\begin{figure}[ht]
\begin{center}
\includegraphics[bb=170 92 490 327,angle=0,scale=0.6]
{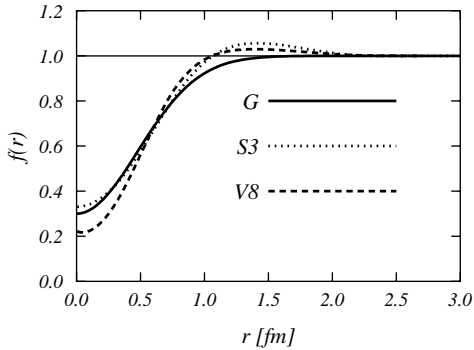}   
\end{center}
\caption{Correlation functions used in our calculations.}
\label{fig:corr}
\end{figure}
\begin{figure}[ht]
\begin{center}
\includegraphics[bb=20 150 600 750,angle=0,scale=0.5]
{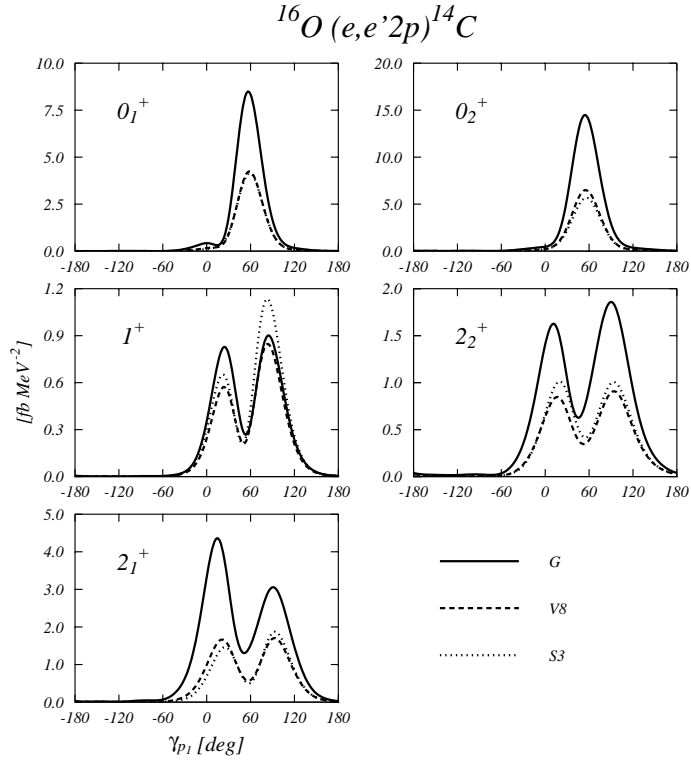}  
\end{center}
\caption{Cross sections for the $^{16}$O(e,e'2p)$^{14}$C
 process. The various curves show the results obtained using the
 correlations of Fig. \protect\ref{fig:corr}.
}
\label{fig:s3v8}
\end{figure}
\begin{figure}[ht]
\begin{center}
\includegraphics[bb=20 150 600 750,angle=0,scale=0.5]
{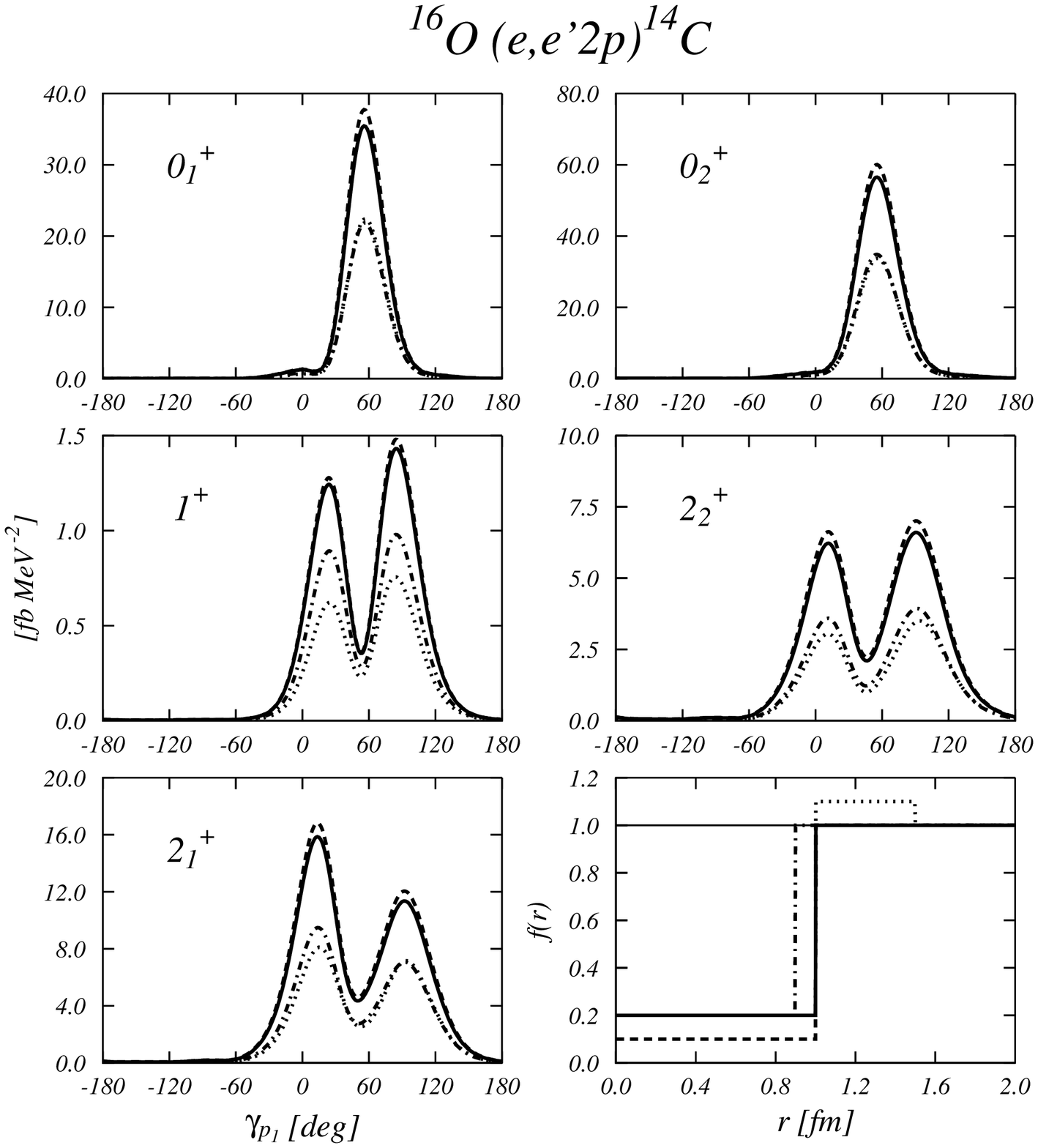}   
\end{center}
\caption{Cross sections for the $^{16}$O(e,e'2p)$^{14}$C
 process calculated with the correlations shown in the lower right
 panel. In this calculations the two-body currents have not been
 considered. 
}
\label{fig:box}
\end{figure}
Two nucleons can be emitted also by two-body electromagnetic currents.
In our case two nucleons of the same type are emitted, therefore the
meson exchange currents where a charged meson is exchanged do not
contribute. There are also two-body currents where a chargeless meson
is exchanged. These currents, implying the excitation of the $\Delta$,
have been considered.

In Fig. \ref{fig:minmax} we show the $^{16}$O(e,e'2p)$^{14}$C cross
sections calculated for various final states. The shaded area is
produced by the uncertainty on the input parameters. These
uncertainties are related to the mean field describing the ground
state, to the optical potential describing the wave functions of the
emitted protons, to the energetics fixing the single particle
energies, and, finally, to the $\gamma$-nucleon-$\Delta$ and the
$\pi^0$-nucleon-$\Delta$ coupling constants.
The uncertainty on the parameters describing the two-body currents
produces the major effect. The 0$^+$ states are rather insensitive to
the two-body currents, while the 1$^+$ state is dominated by the
$\Delta$ excitation even at such relatively low excitation energy.   

Our investigation on the sensitivity of the (e,e'2p) cross section to
the details of the SRC, has been carried on by using the correlation
functions shown in Fig. \ref{fig:corr}.  The G (gaussian) and S3
correlations, have been taken from a Fermi Hypernetted Chain (FHNC)
calculation done with semi-realistic interaction \cite{ari96}.  The V8
correlation is the scalar part of a state dependent correlation used
in FHNC calculation done with a V8' Argonne interaction plus
three-body Urbana IX interaction \cite{fab00}.

The minimum of the G and S3 interaction is almost the same, while the
V8 interaction has a deeper value. The V8 and S3 correlations
overshoot the asymptotic value of 1 in the region between $r$=1 and 2
fm. The cross sections obtained with these correlations are shown in
Fig. \ref{fig:s3v8}.  Apart the result of the 1$^+$ state dominated by
the two-body currents, all the other results have common trends.
First, one should notice that the use of different correlations does
not change the shape of the angular distribution. Second, the cross
sections obtained with the gaussian correlation, are larger than the
other ones.

To understand this result, we have done a set of calculations with
rather schematic correlations. These correlations are shown in the
lower right panel of Fig. \ref{fig:box}.  The cross sections
calculated with these correlations are shown in the other panels by
lines of the same type. In these calculations the two-body $\Delta$
currents have not been included.

The box correlation indicated by the full line is our reference
correlation. Lowering the minimum (dashed lines) does not produce a
large effect. The insertion of a part which overshoots the asymptotic
value reduces the cross section (dotted lines). An analogous effect is
obtained by reducing the size of the box (dashed-dotted lines).

These results can be understood by remembering that the quantity
entering in the cross section calculation is $h(r)=1 - f^2(r)$ .  The
largest is the contribution of $h$ in Eq. (\ref{eq:xi12}), the largest
is the cross section. The overshooting of the asymptotic value,
generates a term in $h(r)$ of opposite sign with respect to the rest
of the function, therefore the total contribution to the integral
becomes smaller. The same effect can be obtained by reducing the size
of the box as it is shown by the dashed dotted lines.

The information about the SRC can be obtained only by a quantitative
comparison between theoretical predictions and experimental data.
Qualitative features, such as the shape of the angular distributions,
are not sensitive to the details of the SRC. Unfortunately a precise
quantitative evaluation of the (e,e'2p) cross sections is linked to
theoretical framework used to calculate it, and to the uncertainties
in the required theoretical input. It appears clear that these kind of
experiments cannot be considered as the ultimate tool able to
pin down exactly the characteristics of the SRC correlations. They
should instead be considered as another, useful and interesting,
element of a puzzle, that together with elastic, inclusive and
one-nucleon emission experiments we are trying to describe in a unique
and coherent theoretical framework.
%
%

%

\begin{thebibliography}{0}
\bibitem{mok00} S.R. Mokhtar, G. Co' and A.M. Lallena,
            {\sl Phys. Rev. C} {\bf 62}, 067304 (2000).
\bibitem{co01} G. Co' and A. M. Lallena, 
            {\it Ann. Phys.} (N.Y.) {\bf 287}, 101 (2001).
\bibitem{ang02} M. Anguiano, G. Co', A. M. Lallena and S.R. Mokhtar, 
            {\sl Ann. Phys.} (N.Y.) {\bf 296}, 235 (2002).
\bibitem{mok01} S. R. Mokhtar, M. Anguiano, G. Co' and A. M. Lallena, 
            {\sl Ann. Phys.} (N.Y.) {\bf 292}, 67 (2001).
\bibitem{ari96} F. Arias de Saavedra, G. Co', A. Fabrocini and S. Fantoni, 
            {\sl Nucl. Phys. A} {\bf 605}, 359 (1996). 
\bibitem{fab00} A. Fabrocini, F. Arias de Saavedra and G. Co',
            {\sl Phys. Rev. C} {\bf 61}, 044302 (2000).
\end{thebibliography}
\end{document}